\documentclass{sig-alternate-05-2015}

\usepackage{lipsum}
\usepackage[numbers,sort]{natbib}
\usepackage{tikz}
\usetikzlibrary{shapes,arrows}
\usepackage{amsmath}
\usepackage{amssymb}
\usepackage{epsfig}
\usepackage[english]{babel}
\usepackage{url}
\usepackage{comment}
\usepackage{balance}
\usepackage{graphicx}
\usepackage{tikz-timing}

\usepackage{multirow}
\usepackage{todonotes}
\usepackage{epstopdf}

\usepackage{array}
\usepackage{subfig}

\usepackage{color}
\usepackage{soul}

\usepackage{enumitem}
\setlist[itemize]{noitemsep,topsep=1pt,parsep=1pt,partopsep=1pt ,leftmargin=0.5cm}%
\setlist[enumerate]{noitemsep,topsep=0pt,parsep=0pt,partopsep=0pt}%

\newcommand{\AppName}{PowerSnitch}

\makeatletter
\def\@copyrightspace{\relax}
\makeatother

\begin{document}
%
\title{No Free Charge Theorem: a Covert Channel via USB Charging Cable on Mobile Devices}

\numberofauthors{6} 
%
\author{
%
%
\alignauthor
Riccardo Spolaor\\
       \affaddr{University of Padua}\\
       \affaddr{Padua, Italy}\\
       \email{rspolaor@math.unipd.it}
\alignauthor
Laila Abudahi\\
\affaddr{University of Washington}\\
\affaddr{Seattle, United States}\\
\email{abudahil@uw.edu} 
\alignauthor 
Veelasha Moonsamy\\
       \affaddr{Radboud University}\\
       \affaddr{Nijmegen, The Netherlands}\\
       \email{veelasha@cs.ru.nl}
\and  
\alignauthor Mauro Conti\\
       \affaddr{University of Padua}\\
       \affaddr{Padua, Italy}\\
       \email{conti@math.unipd.it}
\alignauthor Radha Poovendran\\
       \affaddr{University of Washington}\\
\affaddr{Seattle, United States}\\
\email{rp3@uw.edu} 
}

\maketitle

\begin{abstract}
More and more people are regularly using mobile and battery-powered handsets, such as smartphones and tablets. 
At the same time, thanks to the technological innovation and to the high user demands, those devices are integrating extensive functionalities and developers are writing battery-draining apps, which results in a surge of energy consumption of these devices.
This scenario leads many people to often look for opportunities to charge their devices at public charging stations: the presence of such stations is already prominent around public areas such as hotels, shopping malls, airports, gyms and museums, and is expected to significantly grow in the future. While most of the time the power comes for free, there is no guarantee that the charging station is not maliciously controlled by an adversary, with the intention to exfiltrate data from the devices that are connected to it. 

In this paper, we illustrate for the first time how an adversary could leverage a maliciously controlled charging station to exfiltrate data from the smartphone via a USB charging cable (i.e., without using the data transfer functionality), controlling a simple app running on the device---and without requiring any permission to be granted by the user
to send data out of the device.
We show the feasibility of the proposed attack through a prototype implementation in Android, which is able to send out potentially sensitive information, such as  \texttt{IMEI}, contacts' phone number, and pictures.
\end{abstract}

%

%

\maketitle

\section{Introduction} \label{intro}
Market studies stated that on 2011 smartphone sales have surpassed that of desktop PCs~\cite{p2011}. 
To this date, smartphones remain the most used handheld devices. This is partly due to the fact that these devices are more powerful and provide more functionalities than the traditional feature phones. As a result, users can perform a variety of tasks on an actual smartphone device, which in the past would have been possible only on a desktop PC. 
In order to carry out such tasks, the smartphone platform offers its users a plethora of applications (apps).
 
Moreover, as users are constantly using apps (e.g., the gaming app, Pok\'{e}mon Go) and would eventually require to recharge their smartphones, the demand for public charging stations have increased significantly in the last decade. Such stations can be seen in public areas such as airports, shopping malls, gyms and museums, where users can recharge their devices for free. In fact, this trend is also giving rise to a special type of business\footnote{\url{chargeitspot.com, chargetech.com}}, which allows shop owners to install charging stations in their stores so as to boost their sales by providing free phone recharge to shoppers.

As the phone recharge is for free, however, one cannot be sure that the public charging stations are not maliciously controlled by an adversary. The Snowden revelations gave us proof that civilians are constantly under surveillance and nations are competing against each other by deploying smart technologies for collecting sensitive information en mass. In our work, we consider an adversary (e.g., manufacturers of public charging stations, Government agencies) whose aim is to take control over the public charging station and whose motive is to exfiltrate data from the user's smartphone once the device is plugged into the station.

We demonstrate the feasibility of using power consumption (in the form of power bursts) to send out data over a Universal Serial Bus (USB) charging cable, which acts as a covert channel, to the public charging station. We implemented a proof-of-concept app, \emph{\AppName}, that can send out bits of data in the form of power bursts by manipulating the power consumption of the device's CPU. Interestingly, \AppName~does not require any special permission from the user at install-time (nor at run-time) to exfiltrate data out of the smartphone over the USB cable. On the adversary's side, we designed and implemented a decoder to retrieve the bits that have been transmitted via power bursts. Our empirical results show that we can successfully decode a payload of 512 bits with a 0\% Bit Error Ratio (BER).     
In addition, we stress that the goal of this paper is to assess for the first time the feasibility of data transmission on such a covert channel and not to optimize its performance, which we will tackle as future work. 

We focus primarily on Android as it is currently the leading platform and has a large user base. However, we believe that this attack can be deployed on any other smartphone operating systems, as long as the device is connected to a power source at the public charging station. 

Our contributions are as follows:
\begin{enumerate}

\item To our knowledge, we are the first to demonstrate the practicality of using only the power feature of USB charging cable as a covert channel to exfiltrate data from a device while it is connected to a public charging station.

\item We implemented a prototype of the attack, i.e., we designed and implemented its two components:
\begin{itemize} 
\item
We built a proof-of-concept app, \emph{\AppName}, which does not require any permission granted by the user, 
to communicate bits of information in the form of power bursts back to the adversary.
\item The decoder is deployed on the adversary side, i.e., public charging station to retrieve the binary information embedded in the power bursts.
\end{itemize}
We are able with our prototype to actually send out data using power bursts.
Our prototype demonstrate the practical feasibility of the attack.


\end{enumerate}

The rest of the paper is organized as follows. In {Section~2}, we present a brief literature overview of covert channel and data exfiltration techniques on smartphones. {Section~3} includes some background knowledge on Android and its permission system, and signal transmission and processing. {Section~4} provides a description of our covert channel and decoder design, followed by the experimental results in {Section~5} and discussion in {Section~6}. We conclude the paper in {Section~7}.  

\section{Related Work}

In this section, we survey 
the existing work in the area of covert channels on mobile devices. 
We also present other non-conventional attack vectors, such as side channel information leakage via embedded sensors which can be used for data exfiltration.   

\paragraph{Covert Channels}
A covert channel can be considered as a secret channel used to exfiltrate information from a secured environment in an undetected manner. 
Chandra et al.~\cite{clkk2014} investigated the existence of different covert channels that can be used to communicate between two malicious applications. 
They examined the common resources (such as battery) shared between two malicious applications and how they could be exploited for covert communication. 
Similar studies presented in ~\cite{lw2013,mrfc2012,szzikw2011,nthlz2015} exploited unknown covert channels in malicious and clean applications to leak out private information.   

As demonstrated by Aloraini et al.~\cite{ajsm2015}, the adversary is further empowered as smartphones continue to have more computational power and extensive functionalities. 
The authors empirically showed that speech-like data can be sent over a cellular voice channel. 
The attack was successfully carried out with the help of a custom-built rootkit installed on 
Android devices. 
In~\cite{dmc2015}, Do et al. demonstrated the feasibility of covertly exfiltrating data via SMS and inaudible audio transmission, without the user's knowledge, to other mobile devices including laptops.

In our work, we present a novel covert channel which exploits the USB charging cable by leaking information from a smartphone via power bursts. Our proposed method is non-invasive and can be deployed on non-rooted Android devices. 
We explain the attack in more detail in Section~\ref{subsec:attackOverview}. 

\paragraph{Power Consumption by Smartphones}
In order to prolong the longevity of the smartphone's battery, it is crucial to understand how apps consume  energy during execution and how to optimize such consumption. To this end, several works~\cite{ykjkc2012, pcz2012, ch2010, baghel2012} have been proposed. Furthermore, the authors from~\cite{kss2008, lyzc2009} studied apps' power consumption to detect anomalous behavior on smartphones, thus leading to detection of malware.   

The existing work focus on energy consumption on the device and our attack would therefore go undetected as the smartphone's CPU sends small chunks of encoded data, which is translated into power bursts, back to the public charging station. 

\paragraph{Attack Vectors using Side Channel Leaks}
Modern smartphones are embedded with a plethora of sensors that allow users to interact seamlessly with the apps on their smartphones. 
However, these sensors have access to an abundance of information stored on the device that can get exfiltrated. 
These data leaks can be used as a side channel to infer, otherwise undisclosed, sensitive information about the user or device~\cite{agmbs2010, l2009, ygcm2015}. 

The authors from~\cite{asbm2012, ohdpz2012} demonstrated how accelerometer readings can be used to infer tap-, gesture- and keyboard-based input from users to unlock their smartphones. Similarly, Spreitzer~\cite{s2014} showed that the ambient-light sensor can be exploited to infer users' PIN input by simply observing minor tilts and turns of the smartphones.  

Keystroke inference is another type of attacks that has been successfully demonstrated on the smartphone platform. 
Cai and Chen~\cite{cc2011} used the information collected by motion sensors to deduce the different areas of vibrations on the keypad. 
Maiti et al.~\cite{mjhb2015} applied similar side channel techniques on smartwatches and showed that they can capture individual keystrokes using wrist movements. 
Additionally, the authors from~\cite{mvct2011} proposed a framework that detects and decodes keystrokes by measuring the relative physical position and distance between each vibration. 
Moreover, eavesdropping the network traffic of an Android device, it is possible to identify the set of apps installed on a victim's mobile device~\cite{taylor2016appscanner,stober2013you}, and even infer the actions the victim is performing with a specific app~\cite{conti2016analyzing}.

As pointed out in the aforementioned existing work, the adversarial model did not require any special privileges to exploit side channel leaks to infer data exfiltrated via sensors. 
In this paper, we show that our custom app, \AppName, does not require any special permissions to be granted by the user in order to communicate information (in terms of power bursts) to the adversary.

\section{Background knowledge}
In this section, we briefly recall several concepts that we use in our paper about Android operating system (Section~\ref{android_introduction}), and signal transmission and processing (Section~\ref{signaltransmission}). 

\subsection{Android System and Permissions}
\label{android_introduction} 
\begin{figure}[h]
    \centering
    \includegraphics[width=0.48\textwidth]{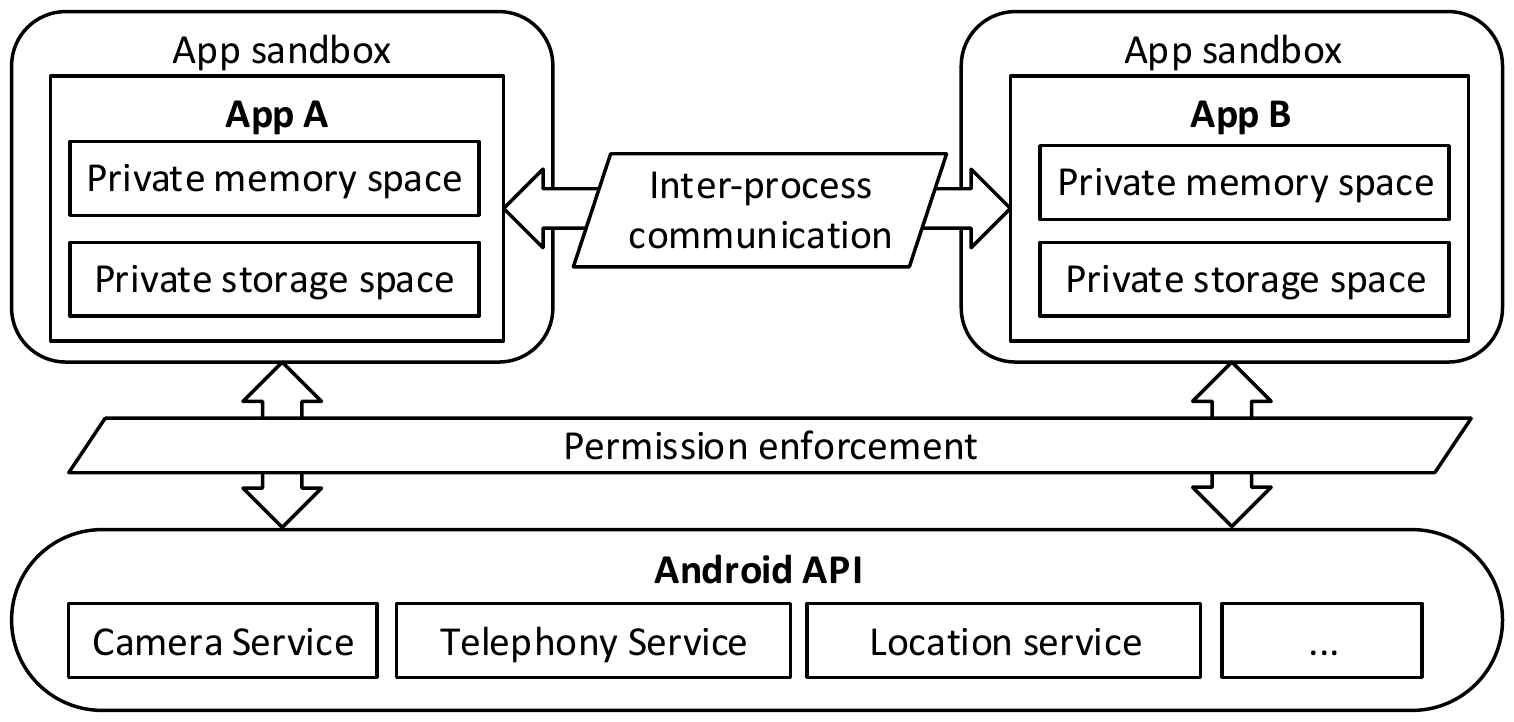}
	\caption{Two sandboxed Android apps and their interaction with one another, and with the Android API.}
	\label{img:android_sandboxing}
\end{figure}

In the Android Operating System (OS), apps are distributed as \texttt{APK} files.
These files are simple archives which contain bytecode, resources and metadata.
A user can install or uninstall an app (thus the APK file) by directly interacting with the smartphone. 
Android apps can be of two kinds:
\begin{itemize}
\item \textit{GUI apps}, which prompt users with a graphical user interface that they can interact with.
\item \textit{Services} that run in the background, independently from user interactions, and provide a service to the user or to other apps.
\end{itemize}
When an Android app is running, its code is executed in a sandbox, as shown in Figure~\ref{img:android_sandboxing}. 
In practice, an app runs isolated from the rest of the system, and it cannot directly access other apps' memory. 
The only way an app could gain memory access is via the mediation of inter-process communication techniques made available by Android.
These measures are in place to prevent the access of malicious apps to other apps' data, which could potentially be privacy-sensitive.

\label{system_apis_and_permissions}
Since Android apps run in a sandbox, they not only have restriction in shared memory usage, but also to most system resources.
Instead, as shown in Figure~\ref{img:android_sandboxing}, the Android OS provides an extensive set of  Accessible Programming Interfaces (APIs), which allows access to system resources 
and services. 
In particular, the APIs that give access to potentially privacy-violating services (e.g., camera, microphone) or sensitive data (e.g., contacts) are protected by the Android Permission System~\cite{AndroidPermissionsDemystified}. 
In fact, an app that wants access to protected data or service must declare the permission (identified by a string) in its manifest file. 
The list of permissions needed by an app is shown to the user when installing the app, and cannot be changed while an app is installed on the device.
With the introduction of Android M (i.e., 6.0), permissions can be dynamically granted 
(by users) during an app's execution. 

The permission system has also been put 
in place in order to reduce the damage due to a successful attack that manages to take control of an app, by limiting the resources that app's process has access to. 
Unfortunately, permission over-provisioning is a common malpractice, so much so that research efforts have been spent in trying to detect this problem~\cite{PermissionGap}.
Moreover, an app asking for permissions not related to its purpose (or functionality) can hide malicious behaviors (i.e., spyware or malware apps)~\cite{mrl2013}.  

Our proof-of-concept app requires the wakelock permission (i.e., \texttt{WAKE\_LOCK}) to wake and force execute the CPU while the device is in sleep mode.
Moreover, since our proposed attack needs also the status of the battery and the USB charging cable, it does not need any permission in order to obtain such information, indeed it is sufficient to only register at run-time (not even in the manifest) a specific broadcast receiver (i.e., \texttt{ACTION\_BATTERY\_CHANGED}).

\newpage
\subsection{Signal Transmission and Processing} 
\label{signaltransmission} 

In this section, we provide some background information on bit transmission, and signal processing and decoding used in our proposed decoder (Section 4.4). 
 
\subsubsection{Bit Transmission }
To enable bit transmission over our channel, an understanding of basic digital communication systems is essential. For proof-of-concept purposes, the design of our bit transmission system was inspired by amplitude-based modulation in the digital communication literature. 

\begin{figure}[!htb]
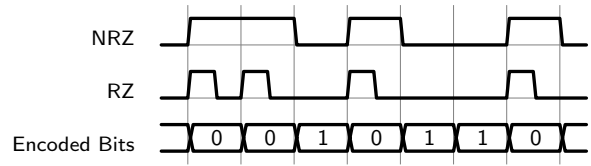

\definecolor{fgred}{rgb}{1 ,0.61569 ,0.61569}%
\resizebox{0.46\textwidth}{!}{
\begin{tikztimingtable}[thick]
 {\tiny NRZ} & LHHHHLLHHLLLLHHL \\\
 {\tiny RZ}  & LHLHLLLHLLLLLHLL \\\
 {\tiny Encoded Bits} & D{} N(B0) 2D{0} N(B4) 2D{0} N(B4) 2D{1} N(B4) 2D{0} N(B4) 2D{1} N(B4) 2D{1} N(B4) 2D{0} N(B4) D{} \\
 \extracode
 
 \begin {pgfonlayer}{background}
 \vertlines [ help lines ]{1 ,3, 5, 7, 9, 11, 13, 15}
 \horlines [ help lines ]{1, 2}
 \end {pgfonlayer}
\end{tikztimingtable}
}
\caption{A comparison between the Non-Return-to-Zero (NRZ) and Return-to-Zero (RZ) On-Off Line Encoding.}
\label{fig:NRZvsRZ}
\end{figure}

Amplitude-Shift Keying (ASK) is a form of digital modulation where digital bits are represented by variations in the amplitude of a carrier signal. To send bits over our channel, we used On-Off Signaling (OOS), which is the simplest form of ASK where digital data is represented by the presence and absence of some pulse $\textit{p(t)}$ for a specific period of time. Figure \ref{fig:NRZvsRZ} shows the difference between a Return-to-Zero (RZ) and a Non-Return-to-Zero (NRZ) on-off encoding. In NRZ encoding, bits are represented by a sufficient condition (a pulse) that occupies the entire bit period T$_{b}$ while RZ encoding represents bits as pulses for a duration of T$_{b}$/2 before it returns to zero for the following T$_{b}$/2 period.  

\begin{figure}[!htb]
\centering
\includegraphics[width=0.45\textwidth]{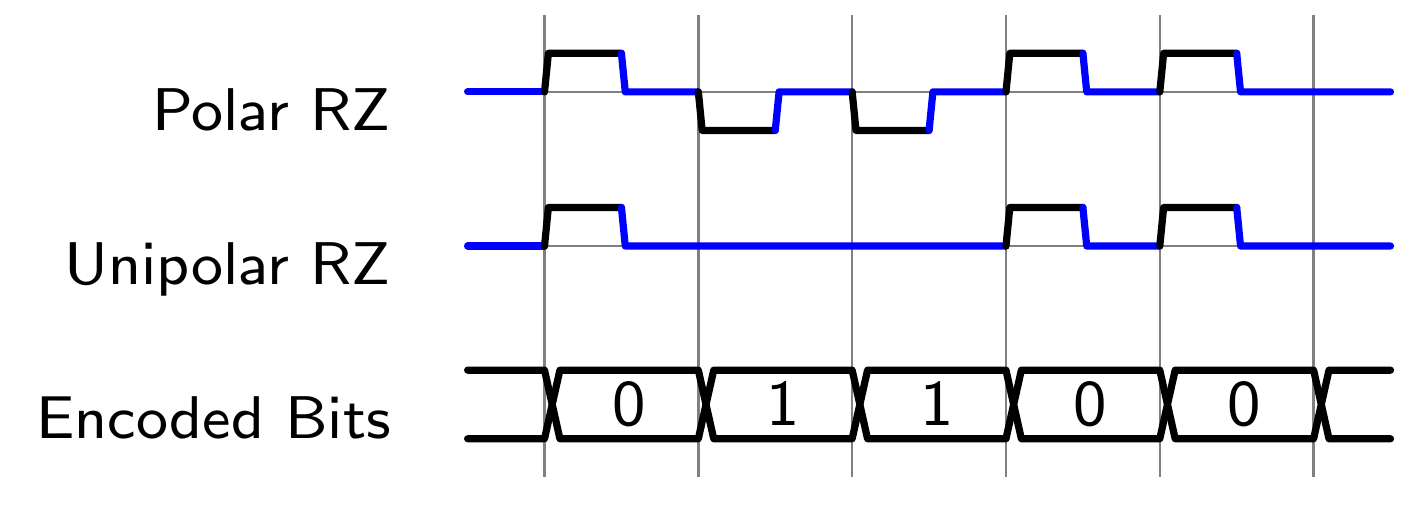}

\caption{A comparison between a Bipolar and a Unipolar encoding of an RZ On-Off Signal.}
\label{fig:polarVSunipolar}
\end{figure}

On the other hand, Figure \ref{fig:polarVSunipolar} shows the difference between a unipolar and a polar RZ on-off signaling. In a polar RZ encoding, two different conditions, different-sign pulses are used to encode different bits(zeros/ones) while the presence and absence of a single pulse, a positive one in our case, are used to encode different bits. \\

For the sake of our channel design, it is safe to assume that we can only increase the power consumption of a phone at certain times and hence, are able to generate only positive (high) bursts. Thus, a unipolar encoding seems more relative and applicable for our channel. Moreover, successive peaks, such as the first two zeros in Figure \ref{fig:NRZvsRZ}, are easier to identify, and thus decode, in the RZ-encoded signal than in the NRZ one. This advantage of RZ over NRZ becomes especially apparent in cases where the bit period is expected not to be restrictively fixed in the received signal whether it is due to expected high channel noises or lack of full control of the phone's CPU. Therefore, unipolar RZ on-off signaling was used to encode leaked bits over our covert channel.

\subsubsection{Signal Processing and Decoding}
After choosing the appropriate encoding method to transmit bits, it is also essential to think about the optimal receiver design and how to process the received signal and decode bits with minimum error probability at the receiver side of the channel. As known in the digital communication literature, matched filters are the optimal receivers for Additive White Gaussian Noise (AWGN) channels. We refer the reader to Section 4.2 of~\cite{Proakis2003} for a detailed proof. 

Matched Filters are obtained by correlating the received signal $R(t)$ with the known pulse that was first used to encode a transmitted bit, in this case P(t) with period T${_b}$. After correlation, the resulted signal is then sampled at time T${_b}$, which means that the sampling rate equals to 1/T${_b}$ samples/seconds. This way, each bit is guaranteed to be represented by only one sample. The decoding decision will then be made based on that one sample value; if the sample value is more than a given threshold, this indicates the presence of P(t); and hence a zero in our case, while a sample value below the threshold indicates the absence of P(t) and hence a one is decoded.

However and most importantly, for matched filters to work as expected, it is essential to have fixed bit period T${_b}$ throughout the entire received signal. If the periods of the received bits were varying, the matched filter samples taken with the 1/T${_b}$ sampling rate will not be as optimal and representative of the bit data as expected and synchronization will be lost. 

Since there exist infrequent phone-specific, OS-enforced conditions that can affect the power consumption of a phone, the noises on our channel are expected to be more complex to fit in an AWGN model. Hence, a matched filter receiver is most likely not the optimal receiver for our channel. More creative decoder design decisions are needed to maximize the throughput of our channel and minimize the error probability. More decoder design decisions and experimental evaluations are presented and discussed in Section 4.4 of the paper.

\section{Covert Channel using Mobile Device Energy Consumption}
In this section, we elaborate on the components that make up our covert channel attack. We begin by giving an overview of the attack in Section~\ref{subsec:attackOverview}. We then define the terms and parameters for transmission in Section~\ref{transprarameters}, followed by a description of each component of the attack: \AppName~ app in Section~\ref{subsec:app} and the energy traces decoder in Section~\ref{decoder1}.

\subsection{Overview of Attack}
\label{subsec:attackOverview}
\begin{figure}[h]
    \centering
    \includegraphics[width=0.48\textwidth]{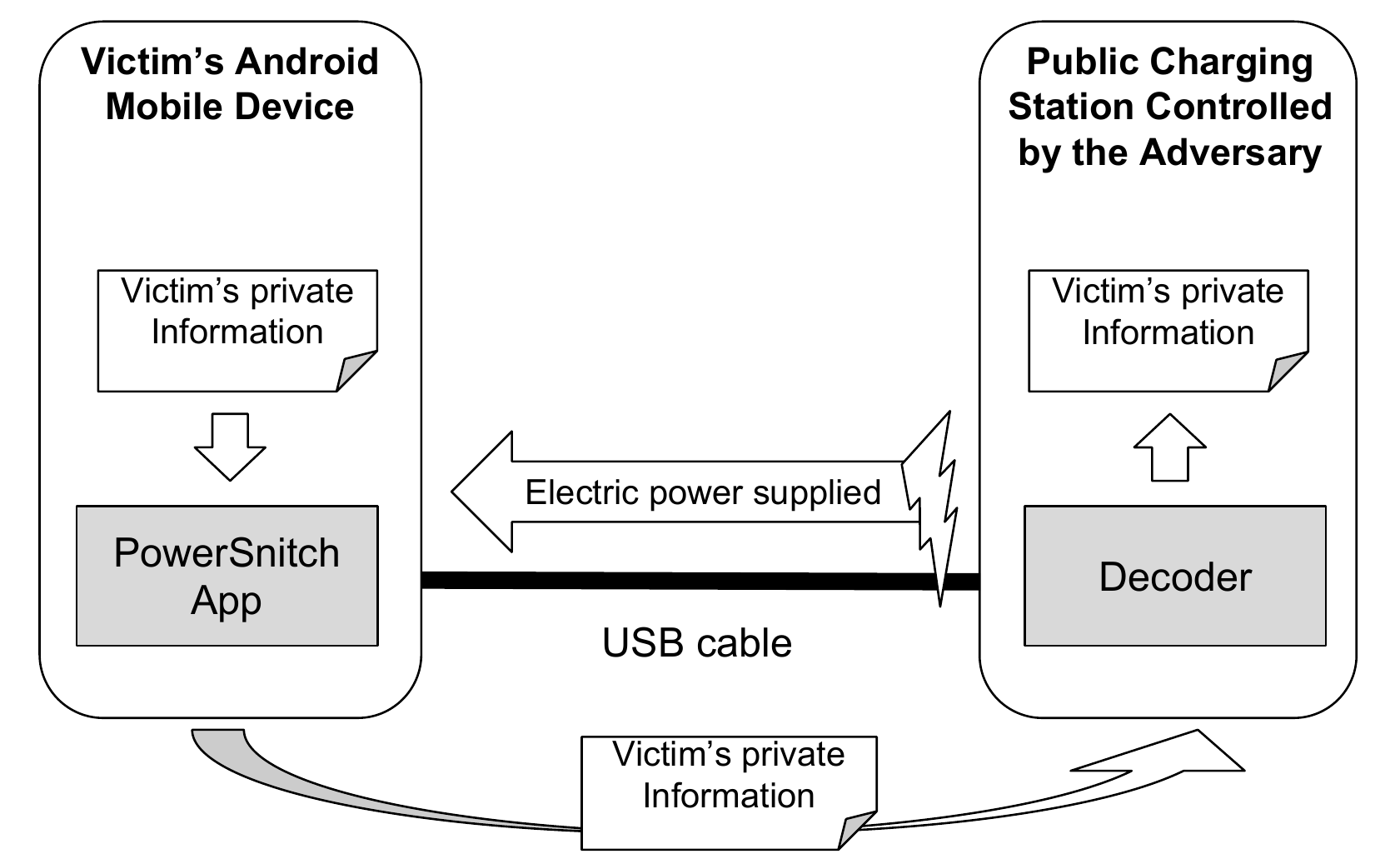}
	\caption{The schema of the components involved in the attack.}
	\label{img:attack_component_schema}
\end{figure}

In this section, we describe the attack scenario.
As reported in Figure~\ref{img:attack_component_schema}, the attack scenario consists of two components: the victim's Android mobile device (sender) and an accomplice power supplier (receiver).
The mobile device is connected to a power supplier through a USB cable in order to recharge its battery.  

The left side of Figure~\ref{img:attack_component_schema} depicts what happens after the victim has installed our proof-of-concept app, \textit{\AppName}. The app is able to exfiltrate the victim's private information, which gets encoded as CPU bursts with a specific timing. 
Indeed, as the CPU is one of the most energy consuming resources in a device, a CPU burst can be directly measured as a peak based on the amount of energy absorbed by a mobile device. The right side of Figure~\ref{img:attack_component_schema} illustrates how the energy supplier is able to measure (with a given sampling rate) the electric current provided to the mobile device connected to the public charging station. 
Then, such electric measurement, which is considered as a \emph{signal}, is given as input to a decoder.


In our proposed covert channel attack, we consider situations in which users connect their mobile devices for more than $20$ minutes. There are several scenarios that fulfill such time requirements, for e.g., recharging a device while waiting at the airport or shopping. In addition, we argue that those time requirements are more than reasonable since generally, 72\% of charging time is more than 30 minutes, with an average time of 3 hours and 54 minutes, as reported in the study of~\cite{ferreira2011understanding}. This means that the mobile device is in stand-by mode and that its CPU and other energy consuming resources (e.g., Wi-Fi or 3/4g data connection) usage is limited only to the OS and background apps.
Moreover, since there is not any user interaction, it is reasonable to assume that the phone screen, which has a relevant impact on energy consumption, will stay off for the aforementioned period of time.

Moreover, it is also worth noting that the attack is still feasible if there is no data connection between the victim's device and the power supplier, such as Media Transfer Protocol (MTP), Photo Transfer Protocol (PTP), Musical Instrument Digital Interface (MIDI). This is possible as our methodology only requires power consumption to send out the power bursts. 
Moreover, from Android version 6.0, when a device is connected via USB, it is set by default to ``Charging'' mode (i.e., just charge the device), thus no data connection is allowed unless the user switches on data connection manually. This improvement in security feature does not impact our proposed attack as we do not make use of data connection to transfer the power bursts.     

\subsection{Terminology and Transmission Parameters}
\label{transprarameters}

In this section, we define the necessary terminology to identify concepts used in the rest of the paper: 
\begin{itemize}
\item \textit{Payload} is the information that has to be sent from the device to the receiver. 
\item \textit{Transmission} is the whole sequence of bits transmitted in which the payload is encoded. 
\end{itemize}

In order to obtain a successful communication, the sender and the receiver need to agree on the parameters of the transmission. 
\begin{itemize}
 \item \textit{Period} is the time interval in which is transmitted a bit. 
 \item \textit{Duty cycle} is the ratio between burst and rest time in a period $T_b$. For example, if a burst lasts for $T_b/2$ the duty cycle will be 50\%. 
 \item \textit{Preamble} is the sequence of bit used to synchronize the transmission. Usually a preamble is used at the beginning of a transmission, but it can also be used within a transmission in order to recover the synchronization in case of error. In our case, we used a preamble composed of 8 bits.
\end{itemize}

\subsection{\AppName~app: Implementing the Attack on Android}
\label{subsec:app}
The first component of our covert channel we discuss is the proof-of-concept which we called \emph{\AppName}.  
This app, used covert channel exploit, has been designed as a service in order to be installed as a standalone app or a library in a repackaged app.
Henceforth, we refer to both these variants simply with the term ``app''.

\begin{figure}[ht!]
\centering
 \subfloat[Permission required.\label{fig:permission}]{%
  \includegraphics[width=0.21\textwidth]{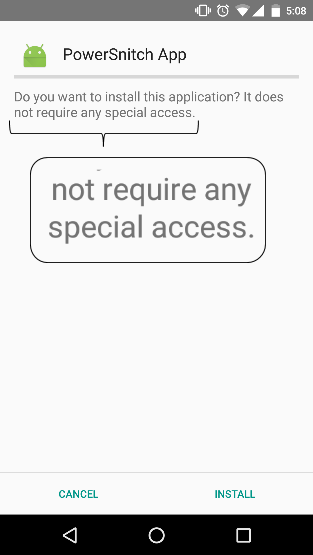}
}
\hfill
 \subfloat[GUI app.\label{fig:mockup}]{%
      \includegraphics[width=0.21\textwidth]{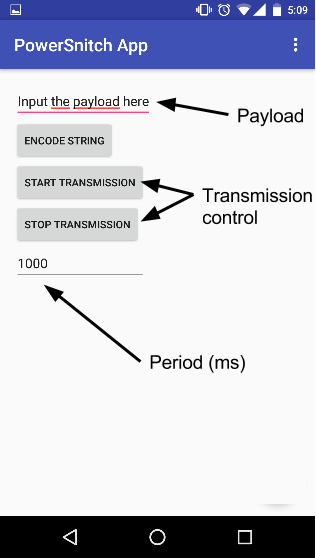}
    }
\caption{Screenshots of the \AppName~in Android 6.0.}
\label{fig:test}
\end{figure}

\AppName~only requires \texttt{WAKE\_LOCK} permission and does not require root access to work. 
The \texttt{WAKE\_LOCK} permission is necessary to wake up the CPU while the phone is in deep sleep mode so that it can start to transmit the payload.
We stress that as it is running as a background service, \AppName~app still works even when user authentication mechanisms (e.g., PIN, password) are in place.
Moreover, since it does not use any conventional communication technology (e.g., Wi-Fi, Bluetooth, NFC), \AppName~app can exfiltrate information even if the device is in airplane mode. 
In this proof-of-concept, we are able to input payload, period and other transmission parameters (see Section~\ref{transprarameters}) to the \AppName~app via an intent from an external app enabled with a Graphic Unit Interface (referred to as GUI app), as depicted in Figure~\ref{fig:test}. 

\begin{figure}[h]
    \centering
    \includegraphics[width=0.48\textwidth]{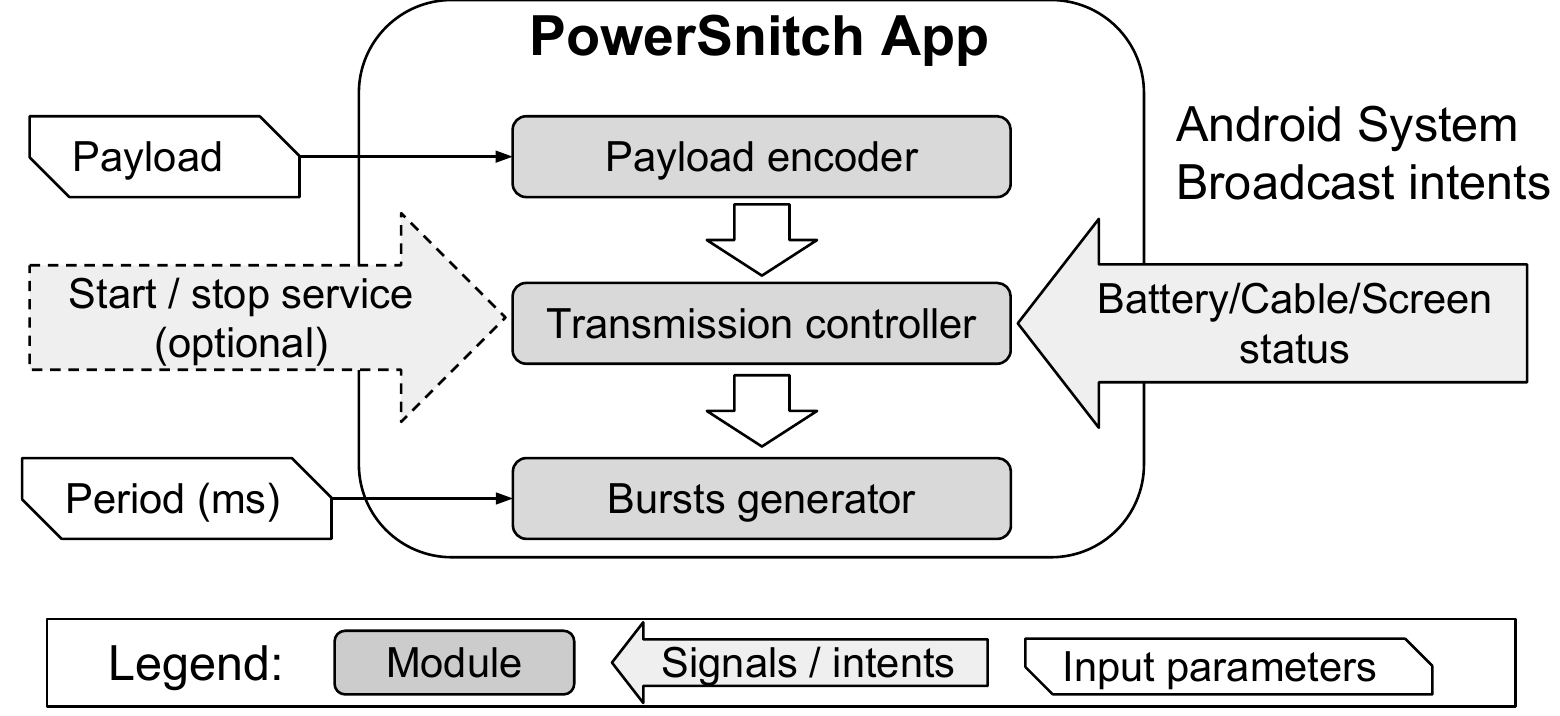}
	\caption{The modules involved in the \AppName~app.}
	\label{img:app_schema}
\end{figure}

In Figure~\ref{img:app_schema}, we illustrate the modules of \AppName~app. It is composed of three modules: the \textit{Payload encoder}, \textit{Transmission controller} and \textit{Bursts generator}. 
The \textit{Payload encoder} module takes the payload that has to be transmitted and gives as output an array of bits. The payload can be any element that can be serialized with an array of bits. 
As a proof-of-concept, we use a string as payload, which is first decomposed into an array of characters and then, using the ASCII code of each character, into an array of bits. 
\textit{Payload Encoder} can also add to its output array synchronization bits (such as the preamble), and for error check and recovery (e.g., CRC).

The \textit{Transmission controller} module is in charge of monitoring the status of the device in order to understand when it is feasible to transmit through the covert channel. 
Indeed, in order to not be detected by the user, it will check whether all the following conditions are satisfied: 
\begin{itemize}
 \item The USB cable has to be connected.
 \item The screen has to be off.
 \item The battery has to be sufficiently charged (see Section~\ref{discussion}).
\end{itemize}
If it receives a broadcast intent from the Android OS that invalidates one of the aforementioned conditions, \textit{Transmission controller} module will interrupt the transmission.
It is worth noticing that to obtain all this information, \AppName~app does not need any additional permission. 
From this GUI app, we are also able to start or stop \AppName~app (represented in Figure~\ref{img:app_schema} with a dotted arrow).

The last component is \textit{Bursts generator} module.
The task of this module is to convert the encoded payload into bursts of energy consumption.
These bursts will generate a signal that can be measured at the other end of the USB cable (i.e., power supplier).
In order to obtain these bursts of energy consumption, \textit{Bursts generator} module can use a power consuming resource of the mobile device such as CPU, screen or flashlight. 
In our proof-of-concept, \textit{Bursts generator} module uses the CPU. A CPU burst is generated from a simple floating point operation repeated in a loop for a precise amount of time (comparing the current timestamp with the one at the start of the loop at each iteration).

\subsection{Analysis of Energy Traces}
\label{decoder1}
To make better decoder design decisions, several channel traces were observed, collected and then used to calculate channel estimations and implement different simulations of the channel performance and behavior. As explained in Section 3, a standard on-off signaling decoder needs to know the exact period of bits in the received signal in order to be able to decode them.
However, a channel built based on a phone's power consumption is expected to have harder-to-model noises that, after examining the collected channel data traces, are actually affecting not only the peak periods but also the peak amplitudes. The amount of external power consumed by a phone can actually be largely affected by dominant OS-enforced, manufacturer-specific factors such as different sudden drop patterns in power consumption especially when the phone is almost or completely charged, lack of control over the OS scheduler; when, how often and for how long do some heavy power-consuming OS background services run, as well as the precision and sampling rate of the power monitor on the receiver side of the channel.

\begin{figure}[!htb]
\centering

       
\includegraphics[width=\columnwidth]{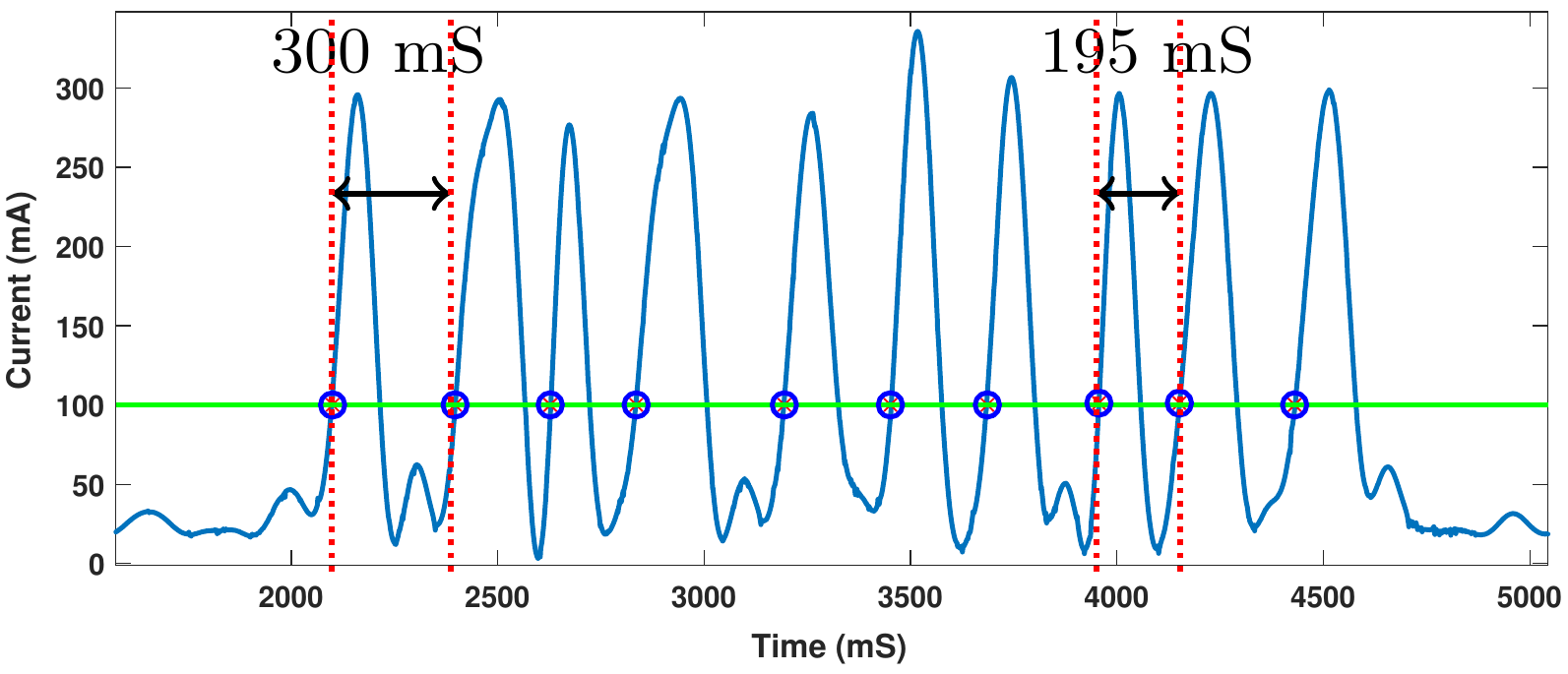} 



\caption{ A portion of a received signal showing the variations in peak widths and amplitudes. }
\label{fig:varying_bitPeriod}
\end{figure}

Figure~\ref{fig:varying_bitPeriod}, for instance, shows a portion of the channel data captured after a transmission of ten successive bits (ten Zeros, therefore ten peaks) was  initiated by our app on a Nexus 6 phone. It should be noted that the data was passed through a low-pass filter to get rid of harsh, high frequency noises in order to make the signal looks smoother. As a result, based on a threshold of $100mA$, ten peaks are successfully detected. Moreover, the width of each peak, and hence the period of each bit, is varying sufficiently. The first bit, for example, has a period of $300ms$ while the eighth one has a period of only $195ms$. Although the intended bit period generated and transmitted by the app was $500ms$, the average period of the received bits was actually $311ms$, which the receiver has no way to predict in advance. Such variations in the received signal are expected to affect the performance of any decoder. As we explained in Section 3.2.2, an ideal matched filter receiver will have hard time decoding such inconsistent signal and synchronization will be lost very quickly. We elaborate further on this issue in the remaining sections.

\subsubsection{Decoder Design}
In this section, we provide additional explanation about the different processing stages that our decoder is taking the received signal through in order to overcome the channel inconsistencies and decode the sent bits with the minimum Bit Error Ratio (BER). In signal processing, the quality of a communication channel can be measured in terms of BER (represented as a percentage), which is the number of bit errors divided by the total number of transmitted bits over the channel. 
Channels affected by interference, distortion, noise, or synchronization errors have a high BER.
 
The processing stages will be discussed in the order they take place in, along with some background information and algorithm justifications, where applicable.

\begin{figure}[!htb]
\centering

 \subfloat[Raw received signal.\label{fig:rawreceived}]{%
  \includegraphics[width=0.49\textwidth]{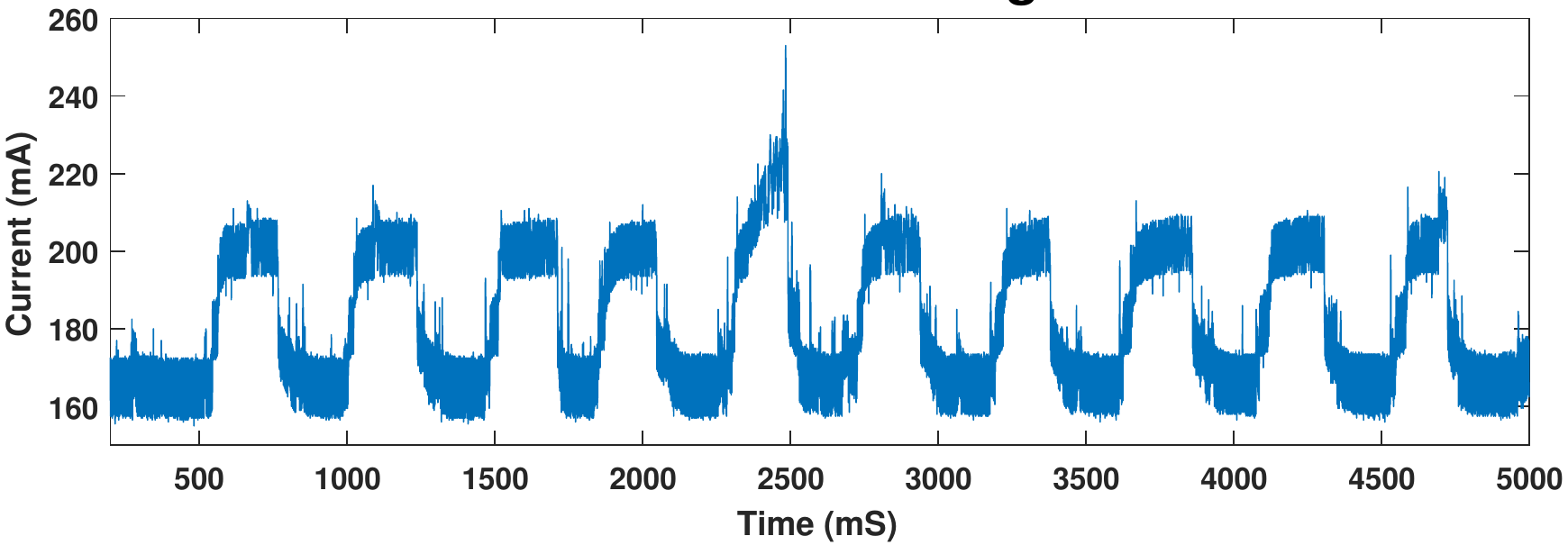}
}
\\
 \subfloat[Low-pass filtered received signal.\label{fig:lowpass1}]{%
      \includegraphics[width=0.49\textwidth]{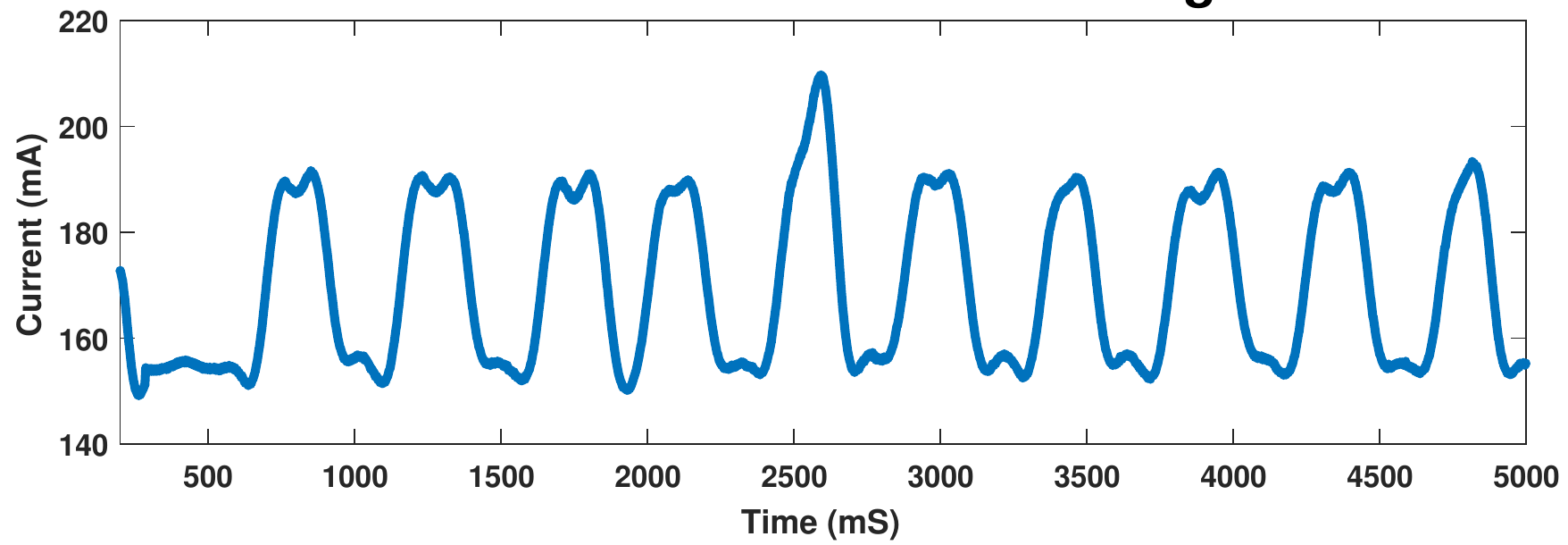}
    }       
\caption{ A portion of a received signal before and after applying the low-pass filter. }
\label{fig:rawvsfiltered}
\end{figure}

\paragraph{Data Filtering} 
First, the received signal is passed through a low-pass filter to get rid of the harsh high-frequency noises. For instance, Figure \ref{fig:rawvsfiltered} shows the same portion of a received signal before and after applying the low-pass filter. The low-pass filter helps not only to make the signal looks smoother, but also to make the threshold-based detection of real peaks easier by eliminating narrow-peak noises that can be falsely identified as real peaks or bits. Additionally, the low-pass filter used in our decoder adjusts its pass and stop frequencies based on the intended bit period generated by the phone in order to make sure that we don't over-filter or over-attenuate the signal. 

\paragraph{Threshold Estimation}
The decoder detects peaks by decoding unipolar RZ on-off encoded bits. The presence or absence of a peak (a $0$ or a $1$ in our case, respectively) at a certain time and for a specific period is then translated to the corresponding bit. Peak detection is usually done by setting an appropriate  threshold; anything above the threshold is a peak and anything below is just noise. However, deciding which threshold to use is not a trivial process especially with the unpredictable noise in our channel and the variations in width and amplitude of the received peaks.  
 
\begin{figure}[ht]
\centering
%
%
 \subfloat[A histogram of the preamble samples.\label{fig:lowpass}]{%
  \includegraphics[width=0.35\textwidth]{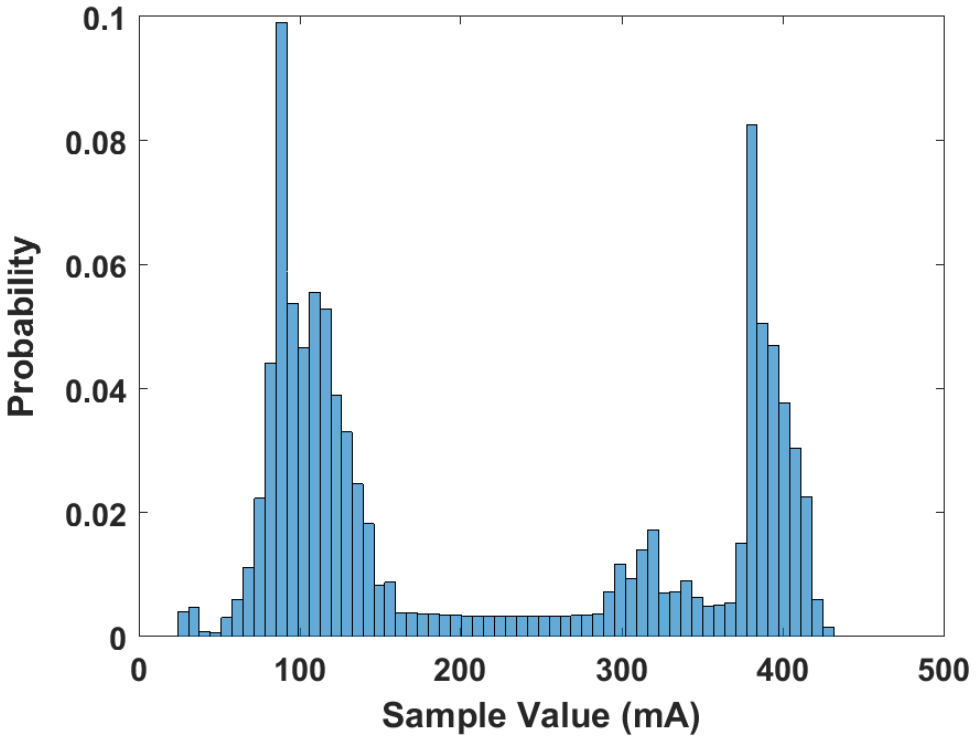}
}
\\
    \subfloat[A received preamble signal.\label{fig:prehist}]{%
     \includegraphics[width=0.35\textwidth]{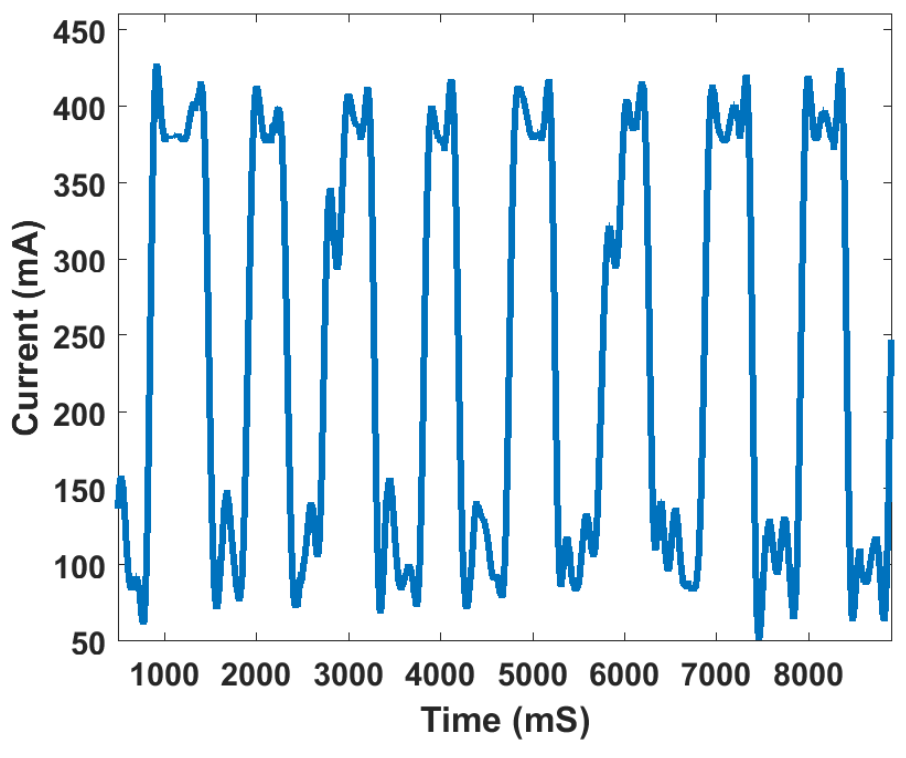}    
}
\caption{A histogram of the preamble samples shows a mixture of two Gaussian-like densities.}
\label{fig:histogram}
\end{figure}

The threshold value used by the decoder is highly critical to peaks detection, the resulted width of detected peaks and the decoder performance. Hence, we primarily use a known preamble data sent prior to the actual packet to estimate the threshold. The preamble consists of eight known bits (eight zeros in our case) at the start of the transmission, which means that the decoder is expecting eight peaks at the start. Since a unipolar RZ on-off encoded zero has a pulse for half of the bit period, the preamble is expected to have roughly the same number of peak and no-peak samples. Therefore, a histogram of the preamble samples is expected to split into two portions; peak and no-peak portions. Figure~\ref{fig:lowpass} shows a histogram of the preamble samples shown in Figure~\ref{fig:prehist}. As observed, the histogram has two distinguishable densities; each of them look like the probability density function of a Gaussian distribution.

\begin{figure}[ht]
\centering
\includegraphics[width=\columnwidth]{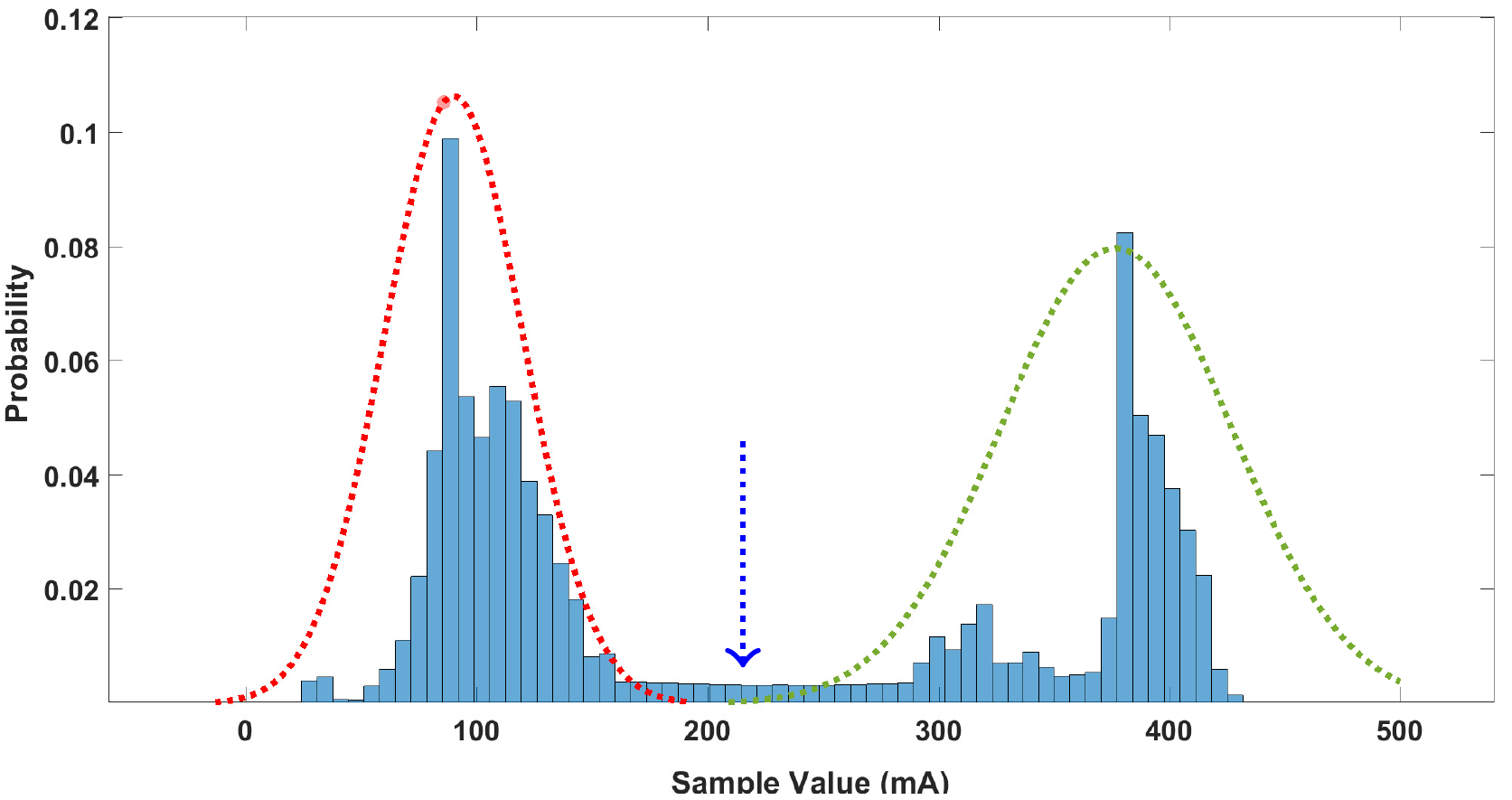}

\caption{ Using the Gaussian Mixture Model to Estimate the Threshold.}
\label{fig:histogramfit}
\end{figure}

Estimating the parameters (mean and variance) of two Gaussians that are believed to exist in one overall distribution is a complicated statistical problem. However, the Gaussian Mixture Model (GMM), introduced and explained in~\cite{reynolds2015gaussian}, is a probabilistic model commonly used to address this type of problem and to statistically estimate the parameters of existing Gaussian populations. To estimate the threshold, as shown in Figure \ref{fig:histogramfit},  the decoder uses the GMM to fit two Gaussians to the two histogram portions, find the mean of each one of them and then compute the threshold as the middle point between the two means. As a result, our decoder is able to estimate the threshold independently and without any previous knowledge of the expected amplitudes of the received bits. After that, each sample is converted to either a peak sample or no-peak sample based on whether the sample value is above or below the estimated threshold. 

\paragraph{Robust Decoding}
Generally, the way a decoder translates the peak and no-peak samples to zeros and ones is highly time-sensitive. For instance, if the bit period is fixed and equals to T$_{b}$, the decoder simply checks the presence or absence of the peak in each T$_{b}$ period. Since this decoding decision is made based on a very strict timing manner, the slightest error in the received bit periods will cause a quick loss of synchronization. As mentioned in the previous section, the received peak widths (and hence bit periods) over our channel are changing with a high variation around their mean. Therefore, our decoding decision cannot rely on an accurate notion of time. Instead, our decoder needs to assume a sufficient amount of error in the period of each received bit and to search for the peaks in a wider range instead of a strict period of time.  

To address this level of time-insensitivity and achieve robustness to synchronization errors, our decoding decision was made based on the time difference between each two successive peaks. As an example, assume that two successive zeros were sent and hence two peaks were received. The difference between the start time of each peak should be rounded to the average bit period. It should be noted that the decoder computes the average bit period based on the received preamble data. However, if a zero-one-zero transmission was made, the time difference of the start of the two received peaks should be rounded to double of the average bit period. If a zero-one-one-zero transmission was made, the difference should be rounded to triple the average period and so on. 
%
Eventually, synchronization is regained with every detected peak and based only on the time difference between peaks, the decoder makes a decision on how many no-peak bits (ones in our case) are transmitted between the zeros. The time difference does not have to be exactly equal to a multiple of the average bit period. Instead, a range of values can be rounded up to the same value and thus more flexible time-insensitive decoding decision is made. 

\section{Experimental Evaluation}\label{sec:experiment}
In this section, we first describe the devices used in our experiments and the values for transmission parameters. We then report the results of the transmission evaluation. 

\subsection{Experiment Settings}
In our experiments, we programmed the \AppName~ app using Android Studio with API. 
The device used to measure the energy provided to the device via USB cable is Monsoon Power Monitor\footnote{\url{www.msoon.com/LabEquipment/PowerMonitor}} in USB mode with $4.55V$ in output.
The decoder used to process signal was implemented in Matlab.
In order to evaluate the performance of the transmission, we send out a payload comprised of letters and numbers of ASCII code for a total of 512 bits.
The values of period used range from $500ms$ to $1000ms$ with increments of $100ms$. It worth mentioning that bits sent over our channel were not packeted and no error detection or correction techniques were used. For each phone and bit period, BER was computed after sending 512 bits at once and then number of bits that were incorrectly decoded was calculated.

We evaluate the performance of our proposal on the following devices running Android OS: Nexus 4 with Android 5.1.1 (API 22), Nexus 5 with Android 6.0 (API 23), Nexus 6 with Android 6.0 (API 23) and Samsung S5 with Android 5.1.1 (API 22). We underline that the devices used in our experiments are actual personal devices, kindly lent by some users without any money reward. In order to replicate an actual real world scenario, we did not uninstalled any app, nor stopped any app running in background.  
The only intervention we made on those devices is the installation of our \AppName~app. 

\subsection{Results}


In Table~\ref{tabresults}, we report the performance of decoder for processing the received power bursts on different mobile devices. 
The results presented in the table are in terms of  Bit Error Ratio (BER) in  the transmission of the payload; the lower the BER, the better is the quality of the transmission.
For Nexus devices (i.e., Nexus 4, 5 and 6), we achieve a zero or low BER of periods of $800ms$ and $900ms$ (i.e., $1.25$ and $1.11$ bits per seconds, respectively).
While for Nexus 4 and 6, the BER remains under 20\%, for Nexus 5, it increases to 37\% and 40\% with periods $700ms$ and $600ms$, respectively. For Samsung S5, the transmission BER is at 12.5\% with a period of 1 second, and it slowly increases to around 21\% with a period of half a second.

\begin{table}[h]
\centering
\small{
\begin{tabular}{ | c | c | c | c | c | c | c |}
\hline
\multirow{2}{*}{\textbf{Device}} & \multicolumn{6}{c|}{\textbf{Period (milliseconds)}}\\ \cline{2-7}
 & 1000 & 900 & 800& 700 & 600 & 500\\
\hline
 
Nexus 4  & 13.5 & 0.78 & 0.0 & 0.0 & 13.33& 16.21\\
Nexus 5  &  21.0& 0.0 & 0.95 & 36.82 & 40.35 & 13.4\\
Nexus 6  &  1.07 & 0.0 & 0.21 & 0.0 & 4.05 &  7.42\\
Samsung S5  & 12.5 & 13.5 & 13.31 & 16.33 & 17.9 & 21.42\\

\hline
\end{tabular}
}
\caption{Results in terms of Bit Error Ratio (BER) as percentage.\label{tabresults}}
\end{table}

The higher BER for Nexus 5 (i.e., periods $700ms$ and $600ms$ in Table~\ref{tabresults}) are due to de-synchronization of the signal that the decoder was not able to recover. 
To cope with this problem, we can divide the payload into packets, where a packet header will be the preamble in order to recover the synchronization. A quick overview of the communication literature can show how a BER of 30\% can be recovered using a simple Forward Error Correction (FEC) technique where the transmitter encodes the data using an Error Correction Code (ECC) prior to transmission; for example bits redundancy or parity checks.


\section{Discussion and optimizations}
\label{discussion}
In this section, we further discuss the results obtained in the experimental evaluation of our proposed attack (Section~\ref{sec:experiment}). In particular, we elaborate on interesting observation we made during our experiments. We also present the optimizations that were implemented in the framework in order to make our proposed attack more robust. 

\begin{figure}[h]
    \centering
    \includegraphics[width=0.46\textwidth]{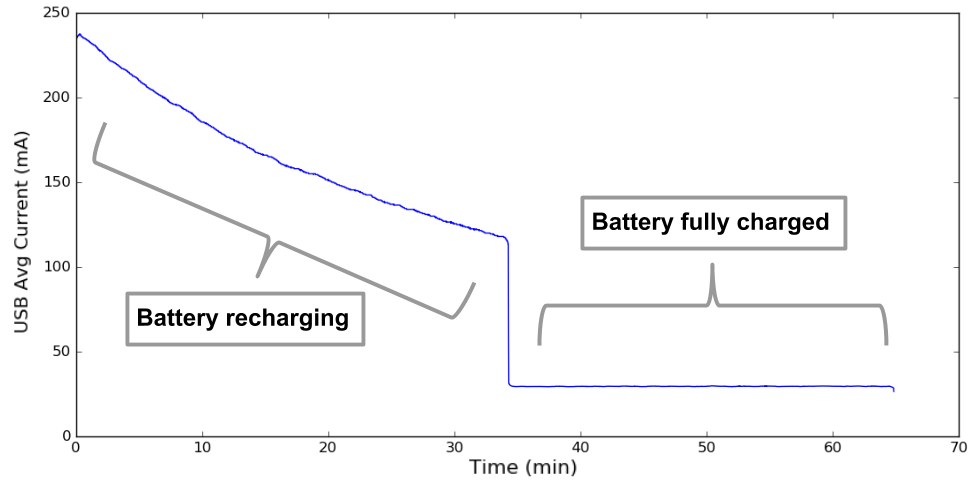} 
	\caption{Electric current provided to a Nexus 6 during recharge phase and battery fully charged.}
	\label{img:rechargen6}
\end{figure}

An interesting thing to notice is that, as observed in our experiments, the level of 
battery affects the quality of the transmission signal.
In Figure~\ref{img:rechargen6}, we present the amount of electric current provided by the power supplier to a Nexus 6 during recharge (i.e., the first 35 minutes) and full battery states (i.e., after 35 minutes). 
Indeed, when the level for the battery is low (i.e., 0\% to around 40\%) the device consumes a high amount of energy, and almost all of it is used to recharge the battery.

When attempting to transmit data in the aforementioned conditions, we discover that the bursts were not easily distinguishable. In fact, the difference in terms of energy consumption between burst and rest was so small that it cannot be distinguished from noise; thus, they can be filtered out during the signal processing. Additionally, when the level of the battery is increased, the amount of energy consumed to recharge the battery gradually decreases. We observed that when the battery level is higher than 50\%, the power bursts become more and more distinguishable. However the best condition under which the bursts are clear is when the battery is fully charged.
Indeed, as we can notice from Figure~\ref{img:rechargen6}, the current drops down after the battery level reaches 100\%, because there is no need to provide energy to the battery anymore - except to keep the device running.

The percentages mentioned above also depends from the power supplier power used to provide energy to the device.
In our experiments, we used Monsoon power monitor which provides in output at most $4.55V$. 
Due to the limitation of such power monitor, during the recharge of devices with fast charge technology (e.g., Samsung S5, Nexus 6 and 6P), which are able to work with $5.3V$ and $2mA$, the energy consumed is almost constant until the battery is almost fully charged. Thus, we cannot decode any signal from the energy consumption.  

In order to avoid to transmit when the receiver is not able to decode the signal, \AppName~checks whether the battery level is among a certain threshold $\omega$.
Such threshold $\omega$ can be obtained by \AppName~itself, simply knowing the model in which it is running. This information can be easily obtained without any permission (\texttt{android.os.Build.MODEL} and \texttt{MANUFACTURER}).

\paragraph{Optimizations} 
In what follows, we elaborate on the optimizations that were implemented in order to not be detected or make the victim suspicious. The first optimization is to keep a duty cycle (i.e., the time of burst in a period) under 50\%. During an attack, if such optimization is not taken into account (i.e., a duty cycle greater than 75\%), the victim may be alerted by two possible effects:
\begin{itemize}
 \item  the temperature of the device could increase significantly, in a way that could be perceived by touching it.
 \item  if the attack takes place during the battery charge phase, the battery will take more time to recharge due to the high amount of energy used by CPU.
\end{itemize}
However, as previously explained in Section~\ref{signaltransmission}, the duty cycle should be 50\% of period (i.e., $T_{b}/2$) in order to achieve a RZ. Thus, the above effects are already taken care of in our proposed attack.  

Another optimization involves the Android Debug Bridge (ADB) tool. It is possible to monitor CPU consumption of an Android device via ADB. Hence, one may use such debug tool to detect that something strange is happening on the device (i.e., a transmission on the covert channel using CPU bursts). Fortunately, \AppName~app could easily detect whether ADB setting is active through \texttt{Settings.Global.ADB\_ENABLED}, once again provided by an Android API. 

Another optimization to \AppName~app would be the ability to detect if the power supplier is an accomplice of the attack.
The accomplice has to let \AppName~app know that it is listening to the covert channel by communicating something equivalent to a ``hello message''.  
In order to do so, we can rely on the information about the amount of electric current provided to recharge the battery. Such information is made available through \texttt{BatteryManager} object, provided by Android API.
In particular, \texttt{BATTERY\_PROPERTY\_CURRENT\_NOW} data field (available from API 21) of \texttt{BatteryManager} records an integer that represents the current entering the battery in terms of mA. 

On one hand, the power supplier can then variate the current in output above and below a certain threshold $\theta$ with a precise timing. 
As a practical and non-limiting example, at a point in time during the recharging, the power supplier can output current with the following behavior: (i) below $\theta$ for $t$ seconds, (ii)above $\theta$ for $t$ seconds, (iii) again below $\theta$ for $t$ seconds and finally (iv) above $\theta$ for good.  
On the other hand, since \AppName~app monitors \texttt{BATTERY\_PROPERTY\_CURRENT\_NOW} and knows the aforementioned behavior (along with both $\theta$ and $t$), it will be able to understand that at the other end of the USB cable there is an accomplice power supplier ready to receive a transmission.
This optimization is significant for reducing the chance to remain undetected, since \AppName~app will transmit data if and only if it is sure that an accomplice power supplier is listening. 
This optimization is not currently implemented, but it considered as future work.

To summarize, the conditions under which the transmission of data is optimal and the chance of being detected is lowest are as follows: the mobile device has to be charged more than 50\%, the screen has to be off, ADB tool should be switched off (which is true by default) and the phone must to be plugged with a USB charging cable to a public charging station which is controlled by the adversary.


\section{Conclusion} \label{conclusion}
In this paper, we demonstrated for the first time the practicality of using a (power-only) USB charging cable as a covert channel to exfiltrate data from a smartphone, which is connected to a public charging station. In order to do so, we implemented an app, \emph{\AppName}, which does not require the user to grant access to permissions at install-time (nor at run-time) on a non-rooted Android phone. Once the device is plugged in a compromised public charging station, the app encodes sensitive information and transmits it via power bursts back to the station. Our empirical results show that we are able to exfiltrate a payload encoded in power bursts at $1.25$ bits per seconds with a BER under 1\% on the Nexus 4-6 devices and a BER of around 13\% for Samsung S5.
As future work, we will work on the transmitter and decoder by extending the framework to include error correction algorithms and synchronization recover mechanisms to lower down the Bit Error Ration of data transmission---as this was not the main goal of this paper.

\section {Acknowledgments}

Veelasha Moonsamy is supported by the Technology Foundation STW (project 13499
- TYPHOON \& ASPASIA) from the Dutch government.

Mauro Conti is supported by a Marie Curie Fellowship funded by the European Commission (agreement PCIG11-GA-2012-321980). This work is also
partially supported by the EU TagItSmart! Project (H2020-ICT30-2015-688061), the EU-India REACH Project (ICI+/2014/342-896), the Italian MIUR-PRIN TENACE Project (20103P34XC), and by the projects ``Tackling Mobile Malware with
Innovative Machine Learning Techniques'', ``Physical-Layer Security for Wireless Communication'', and ``Content Centric Networking: Security and
Privacy Issues'' funded by the University of Padua.



%
%

{
\balance
\bibliographystyle{IEEEtran}
\bibliography{paperbib_slim}

\begin{thebibliography}{10}
\providecommand{\url}[1]{#1}
\csname url@samestyle\endcsname
\providecommand{\newblock}{\relax}
\providecommand{\bibinfo}[2]{#2}
\providecommand{\BIBentrySTDinterwordspacing}{\spaceskip=0pt\relax}
\providecommand{\BIBentryALTinterwordstretchfactor}{4}
\providecommand{\BIBentryALTinterwordspacing}{\spaceskip=\fontdimen2\font plus
\BIBentryALTinterwordstretchfactor\fontdimen3\font minus
  \fontdimen4\font\relax}
\providecommand{\BIBforeignlanguage}[2]{{%
\expandafter\ifx\csname l@#1\endcsname\relax
\typeout{** WARNING: IEEEtran.bst: No hyphenation pattern has been}%
\typeout{** loaded for the language `#1'. Using the pattern for}%
\typeout{** the default language instead.}%
\else
\language=\csname l@#1\endcsname
\fi
#2}}
\providecommand{\BIBdecl}{\relax}
\BIBdecl

\bibitem{p2011}
\textnormal{Business Insider}, ``\textnormal{The Smartphone Market Is Now
  Bigger Than The PC Market},'' 2011, \url{http://goo.gl/XkM8XM}.

\bibitem{clkk2014}
S.~Chandra, Z.~Lin, A.~Kundu, and L.~Khan, ``Towards a systematic study of the
  covert channel attacks in smartphones,'' in \emph{Proc. of SecureComm}, 2014.

\bibitem{lw2013}
J.-F. Lalande and S.~Wendzel, ``Hiding privacy leaks in android applications
  using low-attention raising covert channels,'' in \emph{Proc. of ARES}, 2013.

\bibitem{mrfc2012}
C.~Marforio, H.~Ritzdorf, A.~Francillon, and S.~Capkun, ``Analysis of the
  communication between colluding applications on modern smartphones,'' in
  \emph{Proc. of USENIX ACSAC}, 2012.

\bibitem{szzikw2011}
R.~Schlegel, K.~Zhang, X.~Y. Zhou, M.~Intwala, A.~Kapadia, and X.~Wang,
  ``\textnormal{Soundcomber: A Stealthy and Context-Aware Sound Trojan for
  Smartphones},'' in \emph{Proc. of NDSS}, 2011.

\bibitem{nthlz2015}
E.~Novak, Y.~Tang, Z.~Hao, Q.~Li, and Y.~Zhang, ``Physical media covert
  channels on smart mobile devices,'' in \emph{Proc. of ACM UbiComp}, 2015.

\bibitem{ajsm2015}
B.~Aloraini, D.~Johnson, B.~Stackpole, and S.~Mishra, ``\textnormal{A New
  Covert Channel over Cellular Voice Channel in Smartphones},'' Tech. Rep.,
  2015, arXiv preprint arXiv:1504.05647.

\bibitem{dmc2015}
Q.~Do, B.~Martini, and K.-K.~R. Choo, ``Exfiltrating data from android
  devices,'' \emph{C\&S}, 2015.

\bibitem{ykjkc2012}
C.~Yoon, D.~Kim, W.~Jung, C.~Kang, and H.~Cha, ``\textnormal{AppScope:
  Application Energy Metering Framework for Android Smartphone Using Kernel
  Activity Monitoring},'' in \emph{Proc. of ATC}, 2012.

\bibitem{pcz2012}
A.~Pathak, Y.~Charlie~Hu, and M.~Zhang, ``\textnormal{Where is the energy spent
  inside my app?: fine grained energy accounting on smartphones with Eprof},''
  in \emph{Proc. of ACM EuroSys}, 2012.

\bibitem{ch2010}
A.~Carroll and G.~Heiser, ``An analysis of power consumption in a smartphone,''
  in \emph{Proc. of USENIX ATC}, 2010.

\bibitem{baghel2012}
S.~Baghel, K.~Keshav, and V.~Manepalli, ``An investigation into traffic
  analysis for diverse data applications on smartphones,'' in \emph{Proc. of
  NCC}, 2012.

\bibitem{kss2008}
H.~Kim, J.~Smith, and K.~G. Shin, ``Detecting energy-greedy anomalies and
  mobile malware variants,'' in \emph{Proc. of ACM MobiSys}, 2008.

\bibitem{lyzc2009}
L.~Liu, G.~Yan, X.~Zhang, and S.~Chen, ``\textnormal{VirusMeter: Preventing
  Your Cellphone from Spies},'' in \emph{Proc. of RAID}, 2009.

\bibitem{agmbs2010}
A.~J. Aviv, K.~Gibson, E.~Mossop, M.~Blaze, and J.~M. Smith, ``Smudge attacks
  on smartphone touch screens,'' in \emph{Proc. of USENIX WOOT}, 2010.

\bibitem{l2009}
L.~Lin, T.~Kasper, M.and~G{\"u}neysu, C.~Paar, and W.~Burleson,
  ``\textnormal{Trojan side-channels: lightweight hardware trojans through
  side-channel engineering},'' in \emph{Proc. of CHES}, 2009.

\bibitem{ygcm2015}
L.~Yan, Y.~Guo, X.~Chen, and H.~Mei, ``A study on power side channels on mobile
  devices,'' in \emph{Proc. of ACM Internetware}, 2015.

\bibitem{asbm2012}
A.~J. Aviv, B.~Sapp, M.~Blaze, and J.~M. Smith, ``Practicality of accelerometer
  side channels on smartphones,'' in \emph{Proc. of USENIX ACSAC}, 2012.

\bibitem{ohdpz2012}
E.~Owusu, J.~Han, S.~Das, A.~Perrig, and J.~Zhang, ``\textnormal{ACCessory:
  Password Inference using Accelerometers on Smartphones},'' in \emph{Proc. of
  ACM HotMobile}, 2012.

\bibitem{s2014}
R.~Spreitzer, ``Pin skimming: Exploiting the ambient-light sensor in mobile
  devices,'' in \emph{Proc. of ACM CCS SPSM}, 2014.

\bibitem{cc2011}
L.~Cai and H.~Chen, ``\textnormal{TouchLogger: Inferring Keystrokes on Touch
  Screen from Smartphone Motion},'' in \emph{Proc. of USENIX HotSec}, 2011.

\bibitem{mjhb2015}
A.~Maiti, M.~Jadliwala, J.~He, and I.~Bilogrevic, ``\textnormal{(Smart)watch
  your taps: side-channel keystroke inference attacks using smartwatches},'' in
  \emph{Proc. of ACM ISWC}, 2015.

\bibitem{mvct2011}
P.~Marquardt, A.~Verma, H.~Carter, and P.~Traynor, ``\textnormal{(sp)iPhone:
  decoding vibrations from nearby keyboards using mobile phone
  accelerometers},'' in \emph{Proc. of ACM CCS}, 2011.

\bibitem{taylor2016appscanner}
V.~F. Taylor, R.~Spolaor, M.~Conti, and I.~Martinovic, ``Appscanner: Automatic
  fingerprinting of smartphone apps from encrypted network traffic,'' in
  \emph{Proc. of IEEE EuroS\&P}, 2016.

\bibitem{stober2013you}
T.~St{\"o}ber, M.~Frank, J.~Schmitt, and I.~Martinovic, ``Who do you sync you
  are?: smartphone fingerprinting via application behaviour,'' in \emph{Proc.
  of ACM WiSec}, 2013.

\bibitem{conti2016analyzing}
M.~Conti, L.~V. Mancini, R.~Spolaor, and N.~V. Verde, ``Analyzing android
  encrypted network traffic to identify user actions,'' \emph{IEEE TIFS}, 2016.

\bibitem{AndroidPermissionsDemystified}
A.~P. Felt, E.~Chin, S.~Hanna, D.~Song, and D.~Wagner, ``Android permissions
  demystified,'' in \emph{Proc. of ACM CCS}, 2011.

\bibitem{PermissionGap}
A.~Bartel, J.~Klein, Y.~Le~Traon, and M.~Monperrus, ``Automatically securing
  permission-based software by reducing the attack surface: An application to
  android,'' in \emph{Proc. of ACM ASE}, 2012.

\bibitem{mrl2013}
V.~Moonsamy, J.~Rong, and S.~Liu, ``\textnormal{Mining Permission Patterns for
  Contrasting Clean and Malicious Android Applications},'' \emph{Journal of
  Future Generation Computer Systems}, vol.~36, pp. 122--132, 2013.

\bibitem{Proakis2003}
J.~G. Proakis, \emph{Intersymbol interference in digital communication
  systems}.\hskip 1em plus 0.5em minus 0.4em\relax Wiley Online Library, 2003.

\bibitem{ferreira2011understanding}
D.~Ferreira, A.~K. Dey, and V.~Kostakos, ``Understanding human-smartphone
  concerns: a study of battery life,'' in \emph{Proc. of PerCom}, 2011.

\bibitem{reynolds2015gaussian}
D.~Reynolds, ``Gaussian mixture models,'' \emph{Encyclopedia of biometrics},
  2015.

\end{thebibliography}

}

\end{document}